\documentclass[12pt]{article}
\usepackage{amsmath,amsfonts,amssymb}
\usepackage{amsthm}
\usepackage{graphicx,latexsym}
\usepackage{epsfig}
\usepackage{mathrsfs}
\usepackage{graphicx,epstopdf}
\usepackage{subfigure}
\usepackage{enumitem}
\UseRawInputEncoding
\newcommand{\doublespacing}{\let\CS=\@currsize\renewcommand{\baselinesstrech}
{2.0}\tiny\CS}

\newcommand{\bd}{\begin{document}}
\newcommand{\ed}{\end{document}}
\newcommand{\bc}{\begin{center}}
\newcommand{\ec}{\end{center}}
\newcommand{\bfg}{\begin{figure}}
\newcommand{\efg}{\end{figure}}
\newcommand{\vs}{\vspace}
\newcommand{\beqas}{\begin{eqnarray*}}
\newcommand{\eeqas}{\end{eqnarray*}}
\newcommand{\pad}{\partial}
\newcommand{\f}{\frac}
\DeclareMathOperator{\sech}{sech}

\begin{document}

\title {Characterizing ion-acoustic shock wave collisions in Martian multicomponent plasma environments }

\author{Jayshree Mondal$^1$, Prasanta Chatterjee$^1$, Laxmikanta Mandi$^2$ and Biswajit Sahu$^{3,*}$ \\ $^1$Department of
Mathematics, Siksha Bhavana, Visva Bharati University, \\Santiniketan - 731 235, India \\$^2$Department of Mathematics, University of Gour Banga,
\\Malda 732 103, West Bengal, India\\ $^3$Department of
Mathematics, West Bengal State University, \\Barasat, Kolkata-700 126, India}

  \date{}

  \maketitle
  \vspace{0.5cm}
  \centerline{{\bf ABSTRACT}}

  \vspace {0.2 cm}

  \thispagestyle{empty}

  \setlength{\baselineskip}{18.5 pt}
We present theoretical investigation of colliding ion-acoustic (IA) shock waves in Martian multicomponent plasmas consisting of hydrogen ($H^+$), oxygen ($O^+$) and oxygen molecule ($O_2^+$) ions including background superthermal electrons (modeled by a $\kappa$-(kappa) distribution function). A set of Burgers' equations is obtained adopting a modified Poincar\'e-Lighthill-Kuo (PLK) perturbation method to describe the head-on-collision dynamics of dissipative nonlinear IA wave structures. We have estimated the spatio-temporal scales using parameters typically observed in the Martian atmosphere by the MAVEN spacecraft, for which shock waves are theoretically expected to undergo mutual collisions in the multicomponent plasma. The effects of head-on collisions on the electrostatic potential profiles arising from one-fold and two-fold IA shock interactions are explored. Our numerical analysis reveals that the collision leads to a noticeable broadening of the shock structures with the enhancement of kinematic viscosity.

\vspace {0.7 cm}

\vspace {1.6 cm}
 $\overline{{^* Corresponding ~ author ~ E-mail: biswajit_{-}sahu@yahoo.co.in}}$

\newpage

\section{Introduction}		
A wide range of electrostatic and electromagnetic plasma waves are ubiquitously generated in astrophysical plasma environments, including those near the Earth, other terrestrial planets \cite{tao2012kinetic, malaspina2020plasma, pickett2021review}. These naturally occurring wave modes arise due to the intrinsic presence of multi-component plasmas and play a fundamental role in governing energetic particle dynamics, such as acceleration, deceleration, and transport processes \cite{tesson2022systematic}. Consequently, a comprehensive understanding of the generation mechanisms and nonlinear interactions of these plasma waves is essential for elucidating the complex plasma dynamics in various planetary atmospheres and broader astrophysical systems \cite{saur2018wave, li2021upper, zhao2021mhd}. Numerous observational data from several satellite missions have significantly advanced our understanding of the diverse wave phenomena present in the Earth's magnetospheric environment. Unlike Earth, Mars does not have an internally generated magnetic field. Nevertheless, the interaction between the solar wind and the Martian ionospheric plasma leads to the development of an induced magnetosphere that partially shields the planet from solar wind particles \cite{bertucci2011induced}. The induced magnetosphere of Mars is anticipated to exhibit significant dynamism owing to its direct interaction with the solar wind. Under these conditions, it is likely that a variety of plasma wave phenomena arise within the Martian ionosphere-magnetosphere system. Observations in the Martian upper atmosphere \cite{grard1989first} have revealed the presence of electron plasma oscillations, highlighting the presence of complex plasma wave activity. In November 2013, NASA initiated the Mars Atmosphere and Volatile Evolution (MAVEN) mission to investigate the interaction between the solar wind and the upper atmosphere of Mars \cite{jakosky2015mars}. The MAVEN mission's main objectives are to measure the rate of atmospheric gas loss, comprehend the process causing this loss, examine the interactions between Mars' atmosphere and the solar wind, and gain additional insight into the development of the Martian's atmosphere \cite{jakosky2015mars}. The presence of solitary waves in the Martian magnetosheath has been confirmed through detailed analysis of observational data from the MAVEN mission \cite{kakad2022debye, thaller2022bipolar}.
As per the measurements obtained by MAVEN in the periapsis region at altitudes ranging from 130 to 800 km, it is evident that $H+$, $O+$, and $O_2+$ ions predominantly influence the dayside ionosphere of Mars \cite{akbari2022micro}. The MAVEN mission has significantly advanced our understanding of the Martian plasma environment. MAVEN has revealed key aspects of nonthermal plasma behavior through the identification of diverse ion populations and the detection of superthermal particles \cite{xu2015superthermal, sakai2016electron, fowler2024disappearing}.

In the plasma environment of Mars, a variety of plasma oscillations and waves have been detected, such as whistler waves, proton cyclotron waves, and electron plasma oscillations and waves \cite{yadav2021plasma}. Among these, ion-acoustic waves (IAWs) are considered to be one of the most significant nonlinear waves in plasma. Since, the oscillations of these waves move in the same direction as the wave, they can be referred to as longitudinal waves \cite{jones2012introduction}. IAWs have been observed in the Martian magneto-sheath \cite{kakad2022debye, akbari2022micro}, and the Venusian ionosphere \cite{strangeway1991plasma}, and are also expected at the Martian ionosphere. Recently, Ahmed \textit{et al.} \cite{ahmed2020proliferation} theoretically investigated various nonlinear structures, such as shock waves, explosive solitons, and periodic pulses, in an unmagnetized, collisionless plasma in the lower ionosphere of Titan. Moslem \textit{et al.} \cite{moslem2018shocklike} examined the formation and propagation of shock-like solitons generated by the interaction of solar wind particles with the Venusian ionosphere. Sayed \textit{et al.} \cite{sayed2020nonlinear} explored how IAWs propagate in the Venusian ionosphere by analyzing the effects of relevant plasma parameters on wave behavior. Recently, Varghese \textit{et al.} \cite{varghese2024electrostatic} studied electrostatic solitary waves in electronegative Martian plasma.

However, there are still unresolved aspects of shock waves in the Martian ionosphere, requiring further research to elucidate their underlying mechanisms. One intriguing phenomenon observed within the Martian ionosphere is the occurrence of shock waves, which are abrupt disturbances that can significantly influence plasma dynamics and energy distribution in the region. To the best of our knowledge, no research has been carried out till now to explore the characteristics of IA shock waves in the Martian ionosphere using the data gathered from MAVEN mission \cite{akbari2022micro}. In this study, we focus specifically on the collisions of IA shock waves in the Martian multicomponent plasma environment. The structure of the manuscript is as follows: Section 2 presents a detailed description of the theoretical model, including the key assumptions and the derivation of the coupled Burgers' equations. In Section 3, we obtain analytical solutions to these equations using the Hirota bilinear method. The resulting findings are analyzed and discussed in Sec. 4, while Sec. 5 summarizes the main conclusions of the study.

\section{Basic Equation and Derivation of Both Sided Burgers' Equation}\label{sec2}
We consider a collisionless, unmagnetized, homogeneous, multicomponent plasma with three ionospheric positive ion species $(H^{+},O^{+},$ and $O_2^+)$ and superthermal distributed electron based on the observation of Akbari \textit{et al.} \cite{akbari2022micro}. The dynamics of the nonlinear IAWs is governed by the following basic set of hydrodynamic equations \cite{chen2016plasmas, soltan2025double}

\begin{equation}\label{Eq:1}
\frac{\partial N_{H}}{\partial T}+\frac{\partial N_{H}U_{H}}{\partial X}=0,
\end{equation}

\begin{equation}\label{Eq:2}
\frac{\partial N_{O}}{\partial T}+\frac{\partial N_{O}U_{O}}{\partial X}=0,
\end{equation}

\begin{equation}\label{Eq:3}
\frac{\partial N_{O_2}}{\partial T}+\frac{\partial N_{O_2}U_{O_2}}{\partial X}=0,
\end{equation}

\begin{equation}\label{Eq:4}
\frac{\partial U_{H}}{\partial T}+U_{H}\frac{\partial U_{H}}{\partial X}+\frac{e}{m_{H}}\frac{\partial \phi}{\partial X}+\frac{3K_{B}T_{H}}{m_{H}n_{H0}^2}N_{H}\frac{\partial N_{H}}{\partial X}+\frac{\nu_{1Hi}n_{H0}}{m_{H}N_{H}}\frac{\partial^2 U_{H}}{\partial X^2}=0,
\end{equation}

\begin{equation}\label{Eq:5}
\frac{\partial U_{O}}{\partial T}+U_{O}\frac{\partial U_{O}}{\partial X}+\frac{e}{m_{O}}\frac{\partial \phi}{\partial X}+\frac{3K_{B}T_{O}}{m_{2}n_{O0}^2}N_{O}\frac{\partial N_{O}}{\partial X}+\frac{\nu_{2Oi}n_{O0}}{m_{O}N_{O}}\frac{\partial^2 U_{O}}{\partial X^2}=0,
\end{equation}

\begin{equation}\label{Eq:6}
\frac{\partial U_{O_2}}{\partial T}+U_{O_2}\frac{\partial U_{O_2}}{\partial X}+\frac{e}{m_{O_2}}\frac{\partial \phi}{\partial X}+\frac{3K_{B}T_{O_2}}{m_{O_2}n_{O_20}^2}N_{O_2}\frac{\partial N_{O_2}}{\partial X}+\frac{\nu_{3O_2i}n_{O_20}}{m_{O_2}N_{O_2}}\frac{\partial^2 U_{O_2}}{\partial X^2}=0,
\end{equation}

\begin{equation}\label{Eq:7}
\epsilon_{0}\frac{\partial^2 \phi}{\partial X^2}=e(N_e-N_H-N_O-N_{O_2}),
\end{equation}
where $N_H$, $N_O$ and $ N_{O_2}$ refer to density of  $H^{+}$, $O^{+}$ and $O^{+}_{2}$ respectively. $U_H$, $U_O$ and $ U_{O_2}$ refer to velocity of  $H^{+}$, $O^{+}$ and $O^{+}_{2}$ respectively. $N_e$ is the electron density.

The generaliged Lorentzian distribution \cite{scudder1981survey, vasyliunas1968low, formisano1973solar, saini2009arbitrary} with functional dependence is of the form $f_0(v)\approx \bigg[1+\frac{v^2}{\kappa\theta^2}\bigg]^{-\kappa-1}$, where the spectral index $\kappa$ is a measure of the slope of the energy specturm of the super thermal electron of velocity distribution function.

Here to model the electron distribution, we have used the non-Maxwellian superthermal distribution as \cite{christon1988energy}
\begin{equation*}
f_k(v)=\frac{n_{e0}}{(\pi \kappa \theta^2)^{3/2}}\frac{\Gamma(\kappa+1)}{\Gamma(\kappa-1/2)}\bigg(1-\frac{e\phi}{\kappa m \theta^2}+\frac{v^2}{2\kappa\theta^2}\bigg)^{-\kappa-1},
\end{equation*}
where $\theta^2=\frac{\kappa-3/2}{\kappa}\frac{K_BT_e}{m}$ and $v=\frac{K_BT_e}{m}$, $K_B$ denotes the Boltzmann constant,  $T_e$ is the temperature of electron, $m$ is the mass of electron.
Here the effective thermal speed $\theta$ is only defined for $\kappa>3/2$.
Kappa distribution reduced to Maxwellian distribution for $\kappa\rightarrow \infty$.

To determine electron density, we integrate the Kappa distribution over velocity space, and obtained
\begin{equation}\label{Eq:8}
N_e=n_{e0}\bigg( 1-\frac{e\phi}{K_BT_e(\kappa-3/2)}\bigg)^{-\kappa+1/2}.
\end{equation}

Here we have normalized the above equations by using the following transformations:
\begin{eqnarray*}
x=\frac{X}{\lambda_{Di}}, \hspace{4mm}t=\omega_{Pi}T,~~~u_i=U_i/C_s,~~ (i=H,O,O_2)~~~n_i=\frac{N_i}{n_{i0}},~~ (i=H,O,O_2,e) ,\\
\phi=\frac{e\phi}{K_BT_e},~~~\lambda_{Di}=\big(\frac{\epsilon_{0}K_BT_e}{n_{O_20}e^2}\big)^{1/2} ,~~~~ C_s=\big(\frac{K_BT_e}{m_{O_2}}\big)^{1/2},~~~~ \omega_{Pi}=\frac{C_s}{\lambda_{Di}}.
\end{eqnarray*}

The space and time variables are normalized by the $O_2^{+}$ Debye length
$\lambda_{Di}=\big(\frac{\epsilon_{0}K_BT_e}{n_{O_20}e^2}\big)^{1/2}$ and  $O_2^{+}$ plasma frequency $\omega_{Pi}^{-1}=\big(\frac{n_{O_20}e^2}{\epsilon_0 m_{O_2}}\big)^{-1/2}$ respectively. And the velocity of the ion fluid is normalized by the $O_2^{+}$ IA speed $C_s=\big(\frac{K_BT_e}{m_{O_2}}\big)^{1/2}$.

After normalization, we get the basic equations as:

\begin{equation}\label{Eq:9}
\frac{\partial n_{H}}{\partial t}+\frac{\partial( n_{H}u_{H})}{\partial x}=0,
\end{equation}
\begin{equation}\label{Eq:10}
\frac{\partial n_{O}}{\partial t}+\frac{\partial( n_{O}u_{O})}{\partial x}=0,
\end{equation}
\begin{equation}\label{Eq:11}
\frac{\partial n_{O_2}}{\partial t}+\frac{\partial (n_{O_2}u_{O_2})}{\partial x}=0,
\end{equation}
\begin{equation}\label{Eq:12}
\frac{\partial u_{H}}{\partial t}+u_{H}\frac{\partial u_{H}}{\partial x}+\frac{1}{\mu_{1}}\frac{\partial \phi}{\partial x}+\frac{3\sigma_1}{\mu_{1}}n_{H}\frac{\partial n_{H}}{\partial x}+\nu_{1H}\frac{\partial^2 u_{H}}{\partial x^2}=0,
\end{equation}
\begin{equation}\label{Eq:13}
\frac{\partial u_{O}}{\partial t}+u_{O}\frac{\partial u_{O}}{\partial x}+\frac{1}{\mu_{2}}\frac{\partial \phi}{\partial x}+\frac{3\sigma_{2}}{\mu_{2}}n_{O}\frac{\partial n_{O}}{\partial x}+\nu_{2O}\frac{\partial^2 u_{O}}{\partial x^2}=0,
\end{equation}
\begin{equation}\label{Eq:14}
\frac{\partial u_{O_2}}{\partial t}+u_{O_2}\frac{\partial u_{O_2}}{\partial x}+\frac{\partial \phi}{\partial x}+3\sigma_{3}n_{O_2}\frac{\partial n_{O_2}}{\partial x}+\nu_{3O_2}\frac{\partial^2 u_{O_2}}{\partial x^2}=0,
\end{equation}
\begin{equation}\label{Eq:15}
\frac{\partial^2 \phi}{\partial x^2}=\alpha_e n_e -\alpha_1 n_H - \alpha_2 n_O -n_{O_2},
\end{equation}

\begin{equation}\label{Eq:16}
n_e=\bigg( 1-\frac{\phi}{(\kappa-3/2)}\bigg)^{-\kappa+1/2}.
\end{equation}

where $\mu_1 = \frac{m_H}{m_{O_2}}$ , $\mu_2 = \frac{m_O}{m_{O_2}}$ , $\sigma_1 = \frac{T_H}{T_e}$ , $\sigma_2 = \frac{T_O}{T_e}$ , $\sigma_3 = \frac{T_{O_2}}{T_e}$ , $\alpha_1 = \frac{n_{H0}}{n_{O_20}}$ , $\alpha_2 = \frac{n_{O0}}{n_{O_20}}$ , $\alpha_e = \frac{n_{e0}}{n_{O_20}}$, $\nu_{1H} = \frac{\nu_{1Hi}\omega_{Pi}}{m_Hn_HC_s^2}$ , $\nu_{2O} = \frac{\nu_{2Oi}\omega_{Pi}}{m_On_OC_s^2}$, and $\nu_{3O_2} = \frac{\nu_{3O_2i}\omega_{Pi}}{m_{O_2}n_{O_2}C_s^2}$.

To derive the Burger's equation we initially consider charge neutrality
 i,e.
 \begin{center}
 $\frac{\partial^2 \phi}{\partial x^2}=0$.
\end{center}
We assume that two IAWs are travelling in opposite directions towards each other in the Martian ionosphere. As time goes on , they collide with each other and then depart.  To analyze the effect of the collision an extended Poincar\'e-Lighthill-Kuo (PLK) perturbation method is applied. For using this method, the stretched coordinates are given as \cite{saini2016head}
\begin{align}
\xi &= \epsilon(x-\lambda t)+ \epsilon^2 P_0(\eta, \tau)+\epsilon^3 P_1(\xi,\eta,\tau)+ \cdots,\label{st1}\\
\eta &= \epsilon(x+\lambda t)+ \epsilon^2 Q_0(\xi, \tau)+\epsilon^3 Q_1(\xi,\eta,\tau)+ \cdots,\\
\tau &= \epsilon^3 t.
\end{align}
Here, $\xi$ and $\eta$ represent the trajectories of the IAWs are traveling toward each other, and $P_0(\eta,\tau)$ and $Q_0(\xi,\tau)$ are undefined functions to be deduced later. The parameter $\lambda$ represents the phase velocity of IAW.
Also the dependent variables are expanded as suggested by the extended PLK method as
\begin{align}
n_i &=1+\sum_{k=1}^{\infty}\epsilon^{k+1}n_i^{(k)}, (i=H,O,O_2)\\
u_i &=0+\sum_{k=1}^{\infty}\epsilon^{k+1}u_i^{(k)}, (i=H,O,O_2)\\
\phi &=0+\sum_{k=1}^{\infty}\epsilon^{k+1}\phi^{(k)}.\label{ex1}
\end{align}
Since $\nu_{1H}$, $\nu_{2O}$ and $\nu_{3O_2}$ are very small, then we consider $\nu_{1H}=\epsilon \nu_1$, $\nu_{2O}=\epsilon \nu_2$ and $\nu_{3O_2}=\epsilon \nu_3$.

After substituting the equations(\ref{st1})-(\ref{ex1}) into the equations (\ref{Eq:9})-(\ref{Eq:16}) and equating the coefficients of lower power of $\epsilon$, we obtain the following first order differential equations:
\begin{equation}
\begin{cases}
\lambda \hat{T}n_H^{(1)}+\hat{X}u_H^{(1)} &=0,\\
\lambda \hat{T}n_O^{(1)}+\hat{X}u_O^{(1)} &=0,\\
\lambda \hat{T}n_{O_2}^{(1)}+\hat{X}u_{O_2}^{(1)} &=0,\\
\lambda \hat{T}u_H^{(1)}+\frac{1}{\mu_1}\hat{X}\phi^{(1)}+ \frac{3\sigma_1}{\mu_{1}}\hat{X}n_H^{(1)} &=0,\\
\lambda \hat{T}u_O^{(1)}+\frac{1}{\mu_2}\hat{X}\phi^{(1)}+ \frac{3\sigma_2}{\mu_{2}}\hat{X}n_O^{(1)} &=0,\\
\lambda \hat{T}u_{O_2}^{(1)}+\hat{X}\phi^{(1)}+ 3\sigma_3\hat{X}n_{O_2}^{(1)} &=0,\\
\alpha_e \left(\frac{\kappa-\frac{1}{2}}{\kappa-\frac{3}{2}}\right)\phi^{(1)}-\alpha_1n_H^{(1)}-\alpha_2n_O^{(1)}-n_{O_2}^{(1)} &=0,\label{lb2}
\end{cases}
\end{equation}
where
\begin{align*}
\hat{T}=-\big(\frac{\partial}{\partial \xi}-\frac{\partial}{\partial \eta}\big), \hspace{3mm}\hat{X}=\big(\frac{\partial}{\partial \xi}+\frac{\partial}{\partial \eta}\big).
\end{align*}
After simplifying system ofequation. (\ref{lb2}), we found the following solutions:
\begin{equation}
\begin{cases}
\phi^{(1)} &=\phi_{\xi}^1+\phi_{\eta}^1 , \hspace{3mm} \phi_{\xi}^1=\phi_{\xi}^1(\xi,\tau),\hspace{2mm}
\phi_{\eta}^1=\phi_{\eta}^1(\eta,\tau),\\
n_H^{(1)} &=\frac{1}{\mu_1\lambda^2-3\sigma_1}(\phi_{\xi}^1+\phi_{\eta}^1),\\
n_O^{(1)} &=\frac{1}{\mu_2\lambda^2-3\sigma_2}(\phi_{\xi}^1+\phi_{\eta}^1),\\
n_{O_2}^{(1)} &=\frac{1}{\lambda^2-3\sigma_3}(\phi_{\xi}^1+\phi_{\eta}^1),\\
u_H^{(1)} &=\frac{\lambda}{\mu_1\lambda^2-3\sigma_1}(\phi_{\xi}^1-\phi_{\eta}^1),\\
u_O^{(1)} &=\frac{\lambda}{\mu_2\lambda^2-3\sigma_2}(\phi_{\xi}^1-\phi_{\eta}^1),\\
u_{O_2}^{(1)} &=\frac{\lambda}{\lambda^2-3\sigma_3}(\phi_{\xi}^1-\phi_{\eta}^1).\label{loe2}
\end{cases}
\end{equation}
Here the phase velocity $\lambda$ of IAWs in the Martian ionosphere is obtained by solving the equation
\begin{equation}
\alpha_e\left(\frac{\kappa-\frac{1}{2}}{\kappa-\frac{3}{2}}\right)-\frac{\alpha_1}{\mu_1\lambda^2-3\sigma_1}-
\frac{\alpha_2}{\mu_2\lambda^2-3\sigma_2}-\frac{1}{\lambda^2-3\sigma_3}=0,
\end{equation}
while the undefined functions $\phi_{\xi}^1(\xi,\tau)$ and $\phi_{\eta}^1(\eta,\tau)$ will be obtained from the next order in $\epsilon$. System of equations (\ref{loe2}) imply that, at the higher order we obtained  two waves, one of which $\phi_{\xi}^1(\xi,\tau),$ traveling to the right, and the other $\phi_{\eta}^1(\eta,\tau)$ traveling to the left.\\
At the next order, we obtained the following solutions
\begin{equation}
\begin{cases}
\phi^{(2)} &=\phi_{\xi}^2+\phi_{\eta}^2 , \hspace{3mm} \phi_{\xi}^2=\phi_{\xi}^2(\xi,\tau),\hspace{2mm}
\phi_{\eta}^2=\phi_{\eta}^2(\eta,\tau),\\
n_H^{(2)} &=\frac{1}{\mu_1\lambda^2-3\sigma_1}(\phi_{\xi}^2+\phi_{\eta}^2),\\
n_O^{(2)} &=\frac{1}{\mu_2\lambda^2-3\sigma_2}(\phi_{\xi}^2+\phi_{\eta}^2),\\
n_{O_2}^{(2)} &=\frac{1}{\lambda^2-3\sigma_3}(\phi_{\xi}^2+\phi_{\eta}^2),\\
u_H^{(2)} &=\frac{\lambda}{\mu_1\lambda^2-3\sigma_1}(\phi_{\xi}^2-\phi_{\eta}^2),\\
u_O^{(2)} &=\frac{\lambda}{\mu_2\lambda^2-3\sigma_2}(\phi_{\xi}^2-\phi_{\eta}^2),\\
u_{O_2}^{(2)} &=\frac{\lambda}{\lambda^2-3\sigma_3}(\phi_{\xi}^2-\phi_{\eta}^2).
\end{cases}
\end{equation}
Furthermore, by considering the next higher order of $\epsilon$,  we leads to
\begin{equation}\label{cbg}
   \begin{aligned}
	\frac{4\lambda\alpha_1\mu_{1}}{\mu_1\lambda^2-3\sigma_1}u_H^{(3)}+
	\frac{4\lambda\alpha_2\mu_{2}}{\mu_2\lambda^2-3\sigma_2}u_O^3+
	\frac{4\lambda}{\lambda^2-3\sigma_3}u_{O_2}^3\\
	=	-\int\bigg( A\frac{\partial \phi_{\xi}^1}{\partial \tau}+B\phi_{\xi}^1 \frac{\partial \phi_{\xi}^1}{\partial \xi}+C\frac{\partial^2 \phi_{\xi}^1}{\partial \xi2}\bigg)\text{d}\eta \\
	-\int\bigg( A\frac{\partial \phi_{\eta}^1}{\partial \tau}-B\phi_{\eta}^1 \frac{\partial \phi_{\eta}^1}{\partial \eta}+C\frac{\partial^2 \phi_{\eta}^1}{\partial \eta2}\bigg)\text{d}\xi\\ -\int \int\bigg( D\frac{\partial P_0}{\partial \eta}-E\phi_{\eta}^1\bigg)\frac{\partial\phi_{\xi}^1}{\partial \xi^2}\text{d}\xi \text{d}\eta \\+\int \int \bigg( D\frac{\partial Q_0}{\partial \xi}-E\phi_{\xi}^1\bigg)\frac{\partial\phi_{\eta}^1}{\partial \eta^2}\text{d}\xi \text{d}\eta,
 \end{aligned}
\end{equation}
where
\begin{align*}
A &=\frac{2\lambda\alpha_1\mu_{1}}{P_1^2}+\frac{2\lambda\alpha_2\mu_{2}}{P_2^2}+\frac{2\lambda}{P_3^2},\\
B &=\frac{3\lambda^2\alpha_1\mu_{1}+3\alpha_1\sigma_1}{P_1^3}+\frac{3\lambda^2\alpha_2\mu_{2}+3\alpha_2\sigma_2}{P_2^3}+\frac{3\lambda^2+3\sigma_3}{P_3^3}-\alpha_eK,\\
C &=\frac{\nu_{1}\lambda\alpha_1\mu_{1}}{P_1^2}+\frac{\nu_{2}\lambda\alpha_2\mu_{2}}{P_2^2}+\frac{\nu_{3}\lambda}{P_3^2},\\
D &= \frac{4\lambda^2\alpha_1\mu_{1}}{P_1^2}+\frac{4\lambda^2
	\alpha_2\mu_{2}}{P_2^2}+\frac{4\lambda^2}{P_3^2},\\
E &= \frac{\alpha_1}{P_1^2}+\frac{\alpha_2}{P_2^2}+\frac{1}{P_3^2}+\alpha_eK,\\
P_i&= \mu_i\lambda^2-3\sigma_i,\hspace{4mm}\textsl{where }i=1,2,3,~~~~~\text{and} ~~~~~
K = \frac{\kappa^2-1/4}{(\kappa-3/2)^2}.
\end{align*}
The first (second) term in the right hand side of equation (\ref{cbg}) will be proportional to $\eta(\xi)$ as the integrated function is independent of $\eta(\xi)$. In order to avoid spurious resonance, the first two secular terms of equation (\ref{cbg}) must be eliminated. So we have
\begin{align}
A\frac{\partial \phi_{\xi}^1}{\partial \tau}+B\phi_{\xi}^1 \frac{\partial \phi_{\xi}^1}{\partial \xi}+C\frac{\partial^2 \phi_{\xi}^1}{\partial \xi^2}=0,\label{burg1}\\
A\frac{\partial \phi_{\eta}^1}{\partial \tau}-B\phi_{\eta}^1 \frac{\partial \phi_{\eta}^1}{\partial \eta}+C\frac{\partial^2 \phi_{\eta}^1}{\partial \eta^2}=0.\label{burg2}
\end{align}
The last two terms in equation (\ref{cbg}) are not secular in this order, but they could be secular in the next order. Hence we have \cite{han2008head}
 \begin{align}
 D\frac{\partial P_0}{\partial \eta}-E\phi_{\eta}^1=0,\\
 D\frac{\partial Q_0}{\partial \xi}-E\phi_{\xi}^1=0.
 \end{align}
Equations (\ref{burg1}) and (\ref{burg2}) are the two sided travelling wave Burgers' equations in the reference frames of $\xi$ and $\eta$, respectively.

\section{Solution of Both Sided Burgers' Equation:}\label{sec3}
 The stationary shock wave solutions of the Burgers' equations (\ref{burg1}) and (\ref{burg2}) are obtained, respectively as \cite{chatterjee2022waves, jannat2015ion}:
\begin{eqnarray}
\phi_\xi^1=\phi_m\left(1+\tanh\left(\frac{\xi-U_0\tau}{\Delta_0}\right)\right),\label{0f}\\
\phi_\eta^1=\phi_m\left(-1-\tanh\left(\frac{\eta-U_0\tau}{\Delta_0}\right)\right),\label{0f1}
\end{eqnarray}
where $\phi_m=\frac{AU_0}{B}$, $\Delta_0=\frac{2C}{AU_0}$ and $U_0$  are the amplitude,the width and the speed of the shock waves respectively.

Next, to show the collision of two shock waves we determine another solution of Burgers' equation by using  Hirota bilinear method.
To derive the Hirota bilinear form of equation (\ref{burg1}) via Bell polynomial we introduce  potential fields $v, w$ by setting
\begin{eqnarray}
    \phi_\xi^1=v_\xi, \hspace{2mm}v=\ln{\frac{f}{g}}, \hspace{2mm}w=\ln{fg}, \hspace{2mm}v=w-p, \hspace{2mm} p=2\ln{g}.
    \label{bel1}
\end{eqnarray}

Substituting equation(\ref{bel1}) in equation (\ref{burg1}) and integrating with respect to $\xi$, we have
\begin{align}\label{eq:51}
    E(v,w)=Av_{\tau}+\frac{B}{2}v_{\xi}^2+ Cw_{2\xi}-Cp_{2\xi}=0.
\end{align}
By using Bell polynomial and impose the condition $B=2C$ equation (\ref{eq:51}) can be re-written as combination of binary Bell polynomial $\mathscr{Y}$ expressions as follows
\begin{align}\label{eq:52}
    E(v,w)=A\mathscr{Y}_{\tau}(v,w)+C\mathscr{Y}_{2\xi}(v,w) -C\mathscr{Y}_{2\xi}(v=0,w=p)=0.
\end{align}
Using the  transformation $v=\ln{f/g}, w=\ln{fg} , p=2\ln{g}$   we have the bilinear form of Burgers' equation (\ref{burg1}) are
\begin{align}\label{bilinear1}
\begin{cases}
  (AD_\tau+CD_{\xi}^{2})\{f.g \}=0, \\
  ( CD_{\xi}^{2})\{g.g\}=0,
\end{cases}
 \end{align}
where $C=B/2$ .

$\hspace{6mm}$ To derive the Hirota bilinear form of equation (\ref{burg2}) via Bell polynomial we introduce potential fields $v$ and $w$ by setting
\begin{align}\label{bell2}
    \phi_\eta^1=
   v_\eta, \hspace{2mm}v=\ln{\frac{h}{g}},\hspace{2mm}w=\ln{hg},\hspace{2mm}v=w-p,\hspace{2mm} p=2\ln{g}.
\end{align}
Substituting equation(\ref{bell2}) in equation (\ref{burg2}) and integrating with respect to $\eta$, we have
\begin{align}\label{eq:55}
    E(v,w)=Av_{\tau}-\frac{B}{2}v_{\eta}^2+ Cw_{2\eta}-Cp_{2\eta}=0.
\end{align}
By using Bell polynomial and impose the condition $B=-2C$ equation (\ref{eq:55}) can be re-written as combination of binary Bell polynomial $\mathscr{Y}$ expressions as follows
\begin{align}\label{eq:56}
    E(v,w)=A\mathscr{Y}_{\tau}(v,w)-C\mathscr{Y}_{2\eta}(v,w)-C\mathscr{Y}_{2\eta}(v=0,w=p) =0.
\end{align}
Using the  transformation $v=\ln{h/g},\hspace{2mm} w=\ln{hg},\hspace{2mm} p=2\ln{g}$   we have the bilinear form of Burges'r equation (\ref{burg2}) are
\begin{align}\label{bilinear2}
\begin{cases}
    (AD_\tau-CD_{\eta}^{2})\{h.g\}=0,\\
    CD_{\eta}^{2}\{g.g\}=0,
    \end{cases}
\end{align}
where $C=-B/2$ .

\subsection{For One-Fold solution:} For one-fold soliton of
equation (\ref{bilinear1}) and equation (\ref{bilinear2}) one can take \cite{zuo2011hirota}

$f=1+e^{\theta_{1}}$ , $g=1$ and $h=1+e^{\psi_1}$

where $\theta_1={k_1\xi-\frac{C}{A} k_1^2\tau} +a_1$ and $\psi_1=l_1\eta+\frac{C}{A} l_1^2\tau+b_1$.

Hence \begin{eqnarray}
\phi_\xi^1=k_1\frac{e^{k_1\xi-\frac{C}{A} k_1^2\tau+a_1}}{1+e^{k_1\xi-\frac{C}{A} k_1^2\tau+a_1}},\label{1f}\\
\phi_\eta^1=l_1\frac{e^{l_1\eta+\frac{C}{A} l_1^2\tau+b_1}}{1+e^{l_1\eta+\frac{C}{A} l_1^2\tau+b_1}}.\label{1f1}
\end{eqnarray}
To study the interaction of one-fold shock solution, we consider
\begin{equation}\label{phi1}
\phi^{(1)}=\phi_\xi^1+\phi_\eta^1.
\end{equation}

\subsection{For Two-Fold Solution:}
For two-fold solution ofequation (\ref{bilinear1}) and equation (\ref{bilinear2}) one can take \cite{zuo2011hirota}\\
$f=1+e^{\theta_1}+e^{\theta_2}$ , $g=1$ and $h=1+e^{\psi_1}+e^{\psi_2}$\\
where $\theta_1=k_1\xi-\frac{C}{A} k_1^2\tau+a_1 $, $\theta_2=k_2\xi-\frac{C}{A} k_2^2\tau+a_2$, $\psi_1=l_1\eta+\frac{C}{A}l_1^2\tau+b_1 $ and\\ $\psi_2= l_2\eta+\frac{C}{A}l_2^2\tau+b_2$.\\
Hence \begin{eqnarray}
\phi_\xi^2 &=& \frac{k_1e^{k_1\xi-\frac{C}{A} k_1^2\tau+a_1}+k_2e^{k_2\xi-\frac{C}{A} k_2^2\tau+a_2}}{1+e^{k_1\xi-\frac{C}{A} k_1^2\tau+a_1}+e^{k_2\xi-\frac{C}{A} k_2^2\tau+a_2}},\label{2f}\\
\phi_\eta^2 &=& \frac{l_1e^{l_1\eta+\frac{C}{A}l_1^2\tau+b_1}+l_2e^{l_2\eta+\frac{C}{A}l_2^2\tau+b_2}}{1+e^{l_1\eta+\frac{C}{A}l_1^2\tau+b_1}+e^{l_2\eta+\frac{C}{A}l_2^2\tau+b_2}}.\label{2f2}
\end{eqnarray}
To study the interaction of two-fold shock solution, we consider
\begin{equation}\label{phi2}
\phi^{(1)}=\phi_\xi^1+\phi_\eta^1.
\end{equation}

\section{Result And Discussion:}\label{sec4}
In this investigation, for numerical simulation the parametric values of the plasma parameters are considered based on the MAVEN missions in
Mars' ionosphere \cite{akbari2022micro} as follows:

 \vspace {0.4 cm}

\begin{tabular}{ |p{6cm}|p{5cm}|  }
 \hline
 \multicolumn{2}{|c|}{Typical values of the plasma parameters in the Martian upper atmosphere} \\
 \hline
 Characteristics &  Martian upper atmosphere\\
 \hline
 Electron density($n_e$) & $43$cm$^{-3}$\\
 Temperature of Electron $(T_{e})$ &$60$eV\\
Density of $H^+(n_{H})$ & $5$cm$^{-3}$\\
Temperature of $H^+(T_{H})$ & $5$eV \\
Density of $O^+(n_{O})$ & $8$cm$^{-3}$\\
Temperature of $O^+(T_{O})$ & $2$eV \\
Density of $O_2^+(n_{O_2})$ & $32$cm$^{-3}$\\
Temperature of $O_2^+(T_{O_2})$ & $2$eV \\
Velocity  of $H^+(u_{H})$ & $35$km/s\\
Velocity  of $O^+(u_{O})$ & $5$km/s\\
Velocity  of $O_2^+(u_{O_2})$ & $5$km/s\\
 $K_B$(Boltzmann Constant) & $1.38\times10^{-23}$m$^2$kgs$^{-2}$K$^{-1}$\\
 $\epsilon_0$ & $8.85\times10^{-12}$c$^2$/N.m$^2$\\
 \hline
\end{tabular}

\vspace {0.4 cm}
\begin{figure}[ht!]
	\centering
	\subfigure[]{\includegraphics[width=0.45\linewidth]{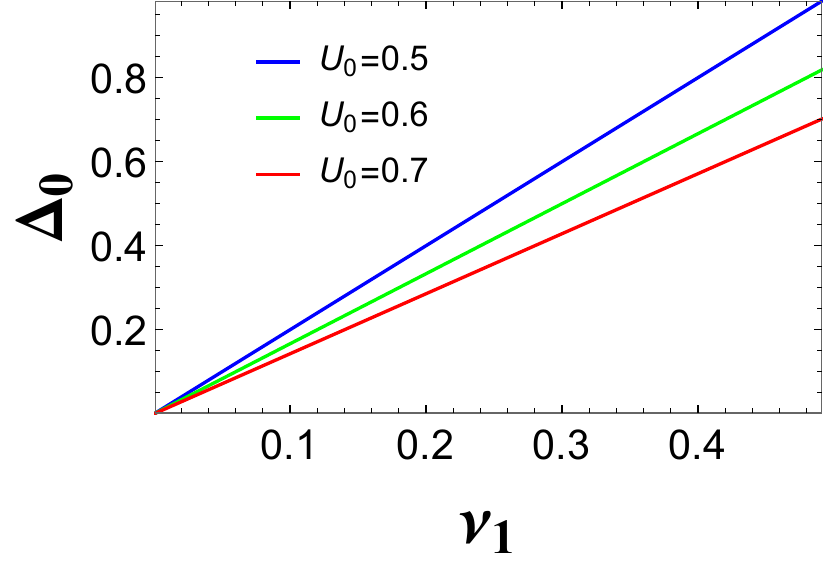}\label{f:0a}}
	\hfill
	\subfigure[]{\includegraphics[width=0.45\linewidth]{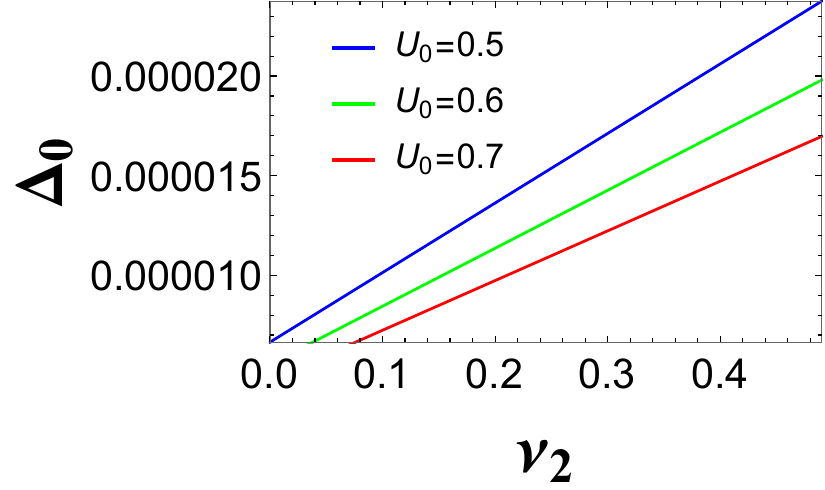}\label{f:0b}}
    \hfill
	\subfigure[]{\includegraphics[width=0.45\linewidth]{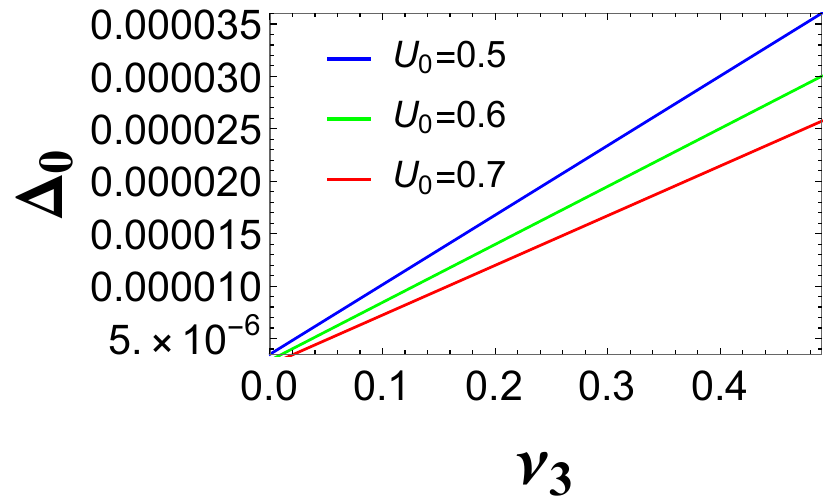}\label{f:0c}}
	\hfill
	\subfigure[]{\includegraphics[width=0.48\linewidth]{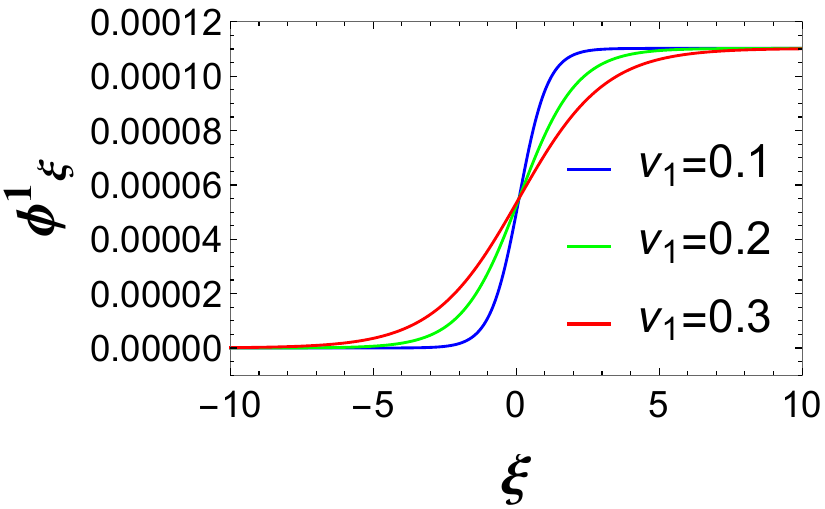}\label{f:0d}}	
	\caption{ (a)-(c) The variation of the shock wave width $\Delta_0$ with viscosities $\nu_1$ (where $\nu_2=0.1$ and $\nu_3=0.1$), $\nu_2$ (where $\nu_1=0$, $\nu_3=0.1$), $\nu_3$ (where $\nu_1=0$, $\nu_2=0.1$), respectively for different values of  $U_0$ and \ref{f:0d} the effect of shock wave  corresponding to equation (\ref{0f}) for different $\nu_1$. The other parameters are $\lambda=2$, $\alpha_1=0.1526$, $\alpha_2=0.25$, $\alpha_e=1.344$, $\mu_1=0.628$, $\mu_2=0.5$, $\sigma_1=0.083$, $\sigma_2=0.033=\sigma_3$, $\kappa=3.5$.}\label{f:0}
\end{figure}

The variation of shock width with shock velocity $U_0$ for different values of viscosity parameters ($\nu_1$, $\nu_2$, $\nu_3$) is presented in Figs. 1(a)-1(c). It is observed that the width of shock structures enhances as the effect of viscosity increases and decreases with increase in group velocity ($U_0$). This behavior can be attributed to the role of molecular viscosity in the system. A higher molecular viscosity enhances the transport of momentum from the nonlinear steepened region of the wave toward the undisturbed leading edge. This efficient redistribution of momentum acts to smooth out sharp gradients, resulting in a wider shock front. The development of such shock structures is governed by a sensitive interplay between nonlinear steepening effects, which tend to create sharper features, and dissipative mechanisms, such as viscosity, which counteract steepening by spreading the wave energy. Figure 1(d)  illustrates the variation of positive potential shock profile for different values of viscosity parameter $\nu_1$. We found that the amplitude of IA shock waves remains unaffected by variations in the dissipation parameter. However, the width of the shock wave structures exhibits a direct dependence on $\nu_1$, increasing progressively as the dissipative coefficient becomes larger. This indicates that while dissipation does not alter the energy concentration at the crest of the wave, it significantly influences the spatial spread of the shock front by enhancing the smoothing effect across the wave profile.
\begin{figure}[ht!]
	\centering
	\subfigure[]{\includegraphics[width=0.45\linewidth]{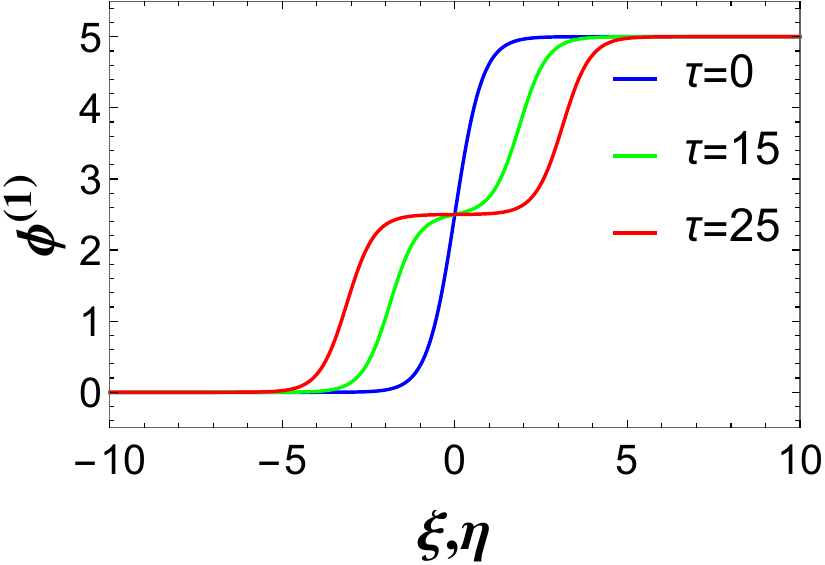}\label{f:1a}}
	\hfill
	\subfigure[]{\includegraphics[width=0.45\linewidth]{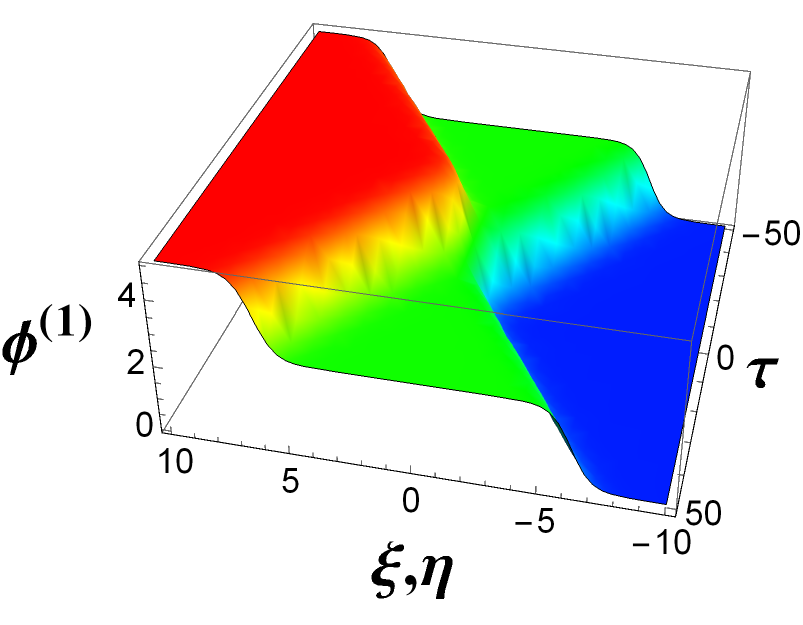}\label{f:1b}}
	\hfill
	\subfigure[]{\includegraphics[width=0.45\linewidth]{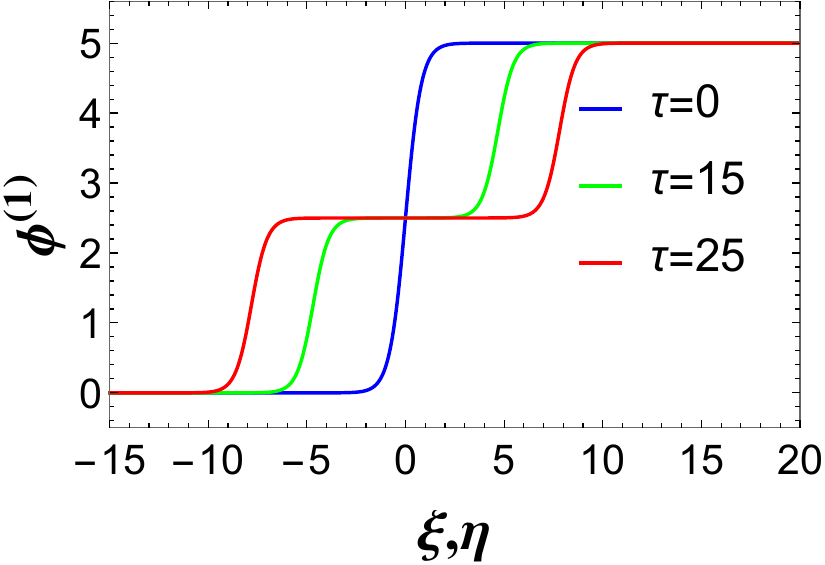}\label{f:1c}}
	\hfill
	\subfigure[]{\includegraphics[width=0.45\linewidth]{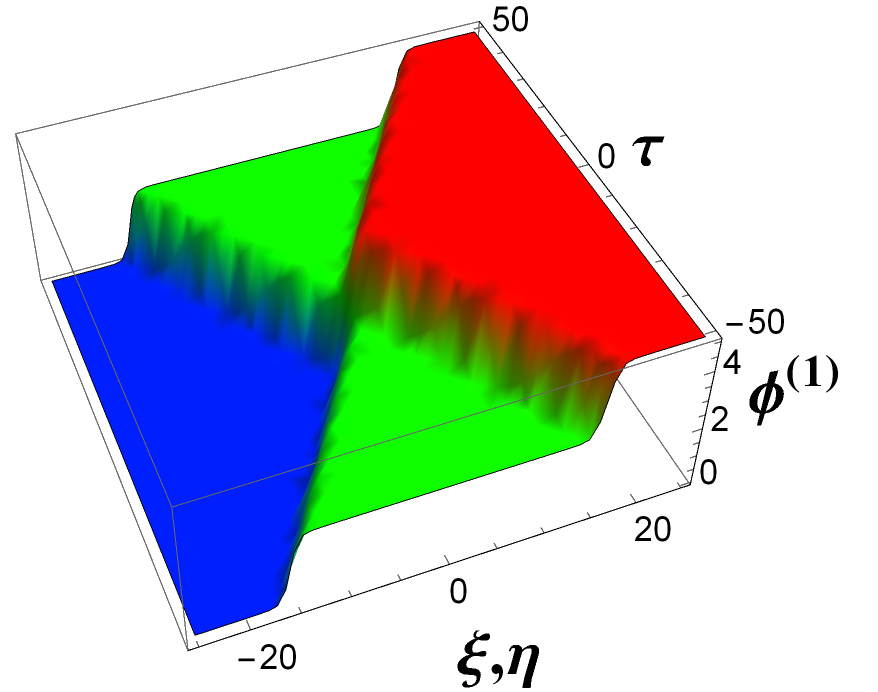}\label{f:1d}}
	\caption{ Collision of one-fold shock wave corresponding to equation (\ref{phi1}) for $\lambda=2$, $\alpha_1=0.1526$, $\alpha_2=0.25$, $\alpha_e=1.344$, $\mu_1=0.628$, $\mu_2=0.5$, $\sigma_1=0.083$, $\sigma_2=0.033=\sigma_3$,$\kappa=3.5$, $\nu_2=0.0$, $\nu_3=0.1$ and $k_1=2.5=l_1$, where (a) and (c) are profile for different $\tau$ where $\nu_1=0.1$ and $\nu_1=0.25$ respectively and (b), (d) are corresponding 3D figure of (a) and (c), respectively}.\label{f:1}
\end{figure}

Figure 2 shows the temporal evolution of interaction of two one-fold shock waves in the Martian multicomponent plasma environments to study the effects of kinematic viscosity for $H^+$. Figures 2(a) and 2(c) illustrate the electrostatic potential profiles resulting from the head-on interaction of shock waves for two different values of the kinematic viscosity parameter $\nu_1$ (specifically, $0.1$ and $0.25$). The corresponding three-dimensional structures of these interactions are exhibited in Figs. 2(b) and 2(d), respectively. It is evident that as the value of $\nu_1$ (the kinematic viscosity for $H^+$) increases, the spatial extent or width of the interacting shock waves becomes broader as time increases. Physically, this broadening effect is a direct consequence of the increased viscous dissipation in the plasma. This behavior arises due to the enhanced momentum diffusion associated with higher viscosity in the fluid. An increase in $\nu_1$ corresponds to a greater resistance to flow deformation, which suppresses the formation of sharp gradients and leads to smoother, more extended shock structures. Thus, the system exhibits a stronger dissipative response, diminishing the steepness of the shocks and enlarging their spatial interaction region. Figures 3(a) and 3(c) illustrate the time evolution of two-fold IA shock wave collisions for two distinct values of kinematic viscosity, namely $\nu_1 = 0.2$ and $\nu_1 = 0.4$, respectively. The associated three-dimensional visualizations are presented in Figs. 3(b) and 3(d), providing a clear perspective on the spatiotemporal dynamics of the shock interactions. It is evident from these figures that increasing the viscosity parameter leads to a notable broadening in the spatial extent of the shock structures during the collision process.

\begin{figure}[ht!]
	\centering
	\subfigure[]{\includegraphics[width=0.45\linewidth]{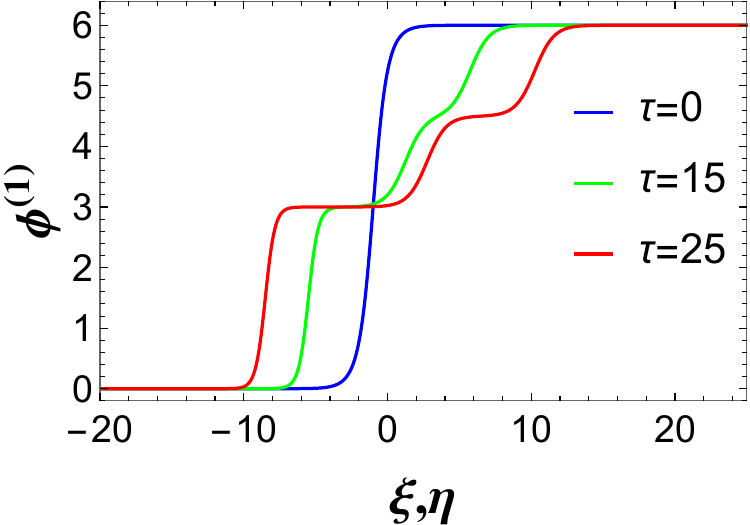}\label{f:2a}}
	\hfill
	\subfigure[]{\includegraphics[width=0.45\linewidth]{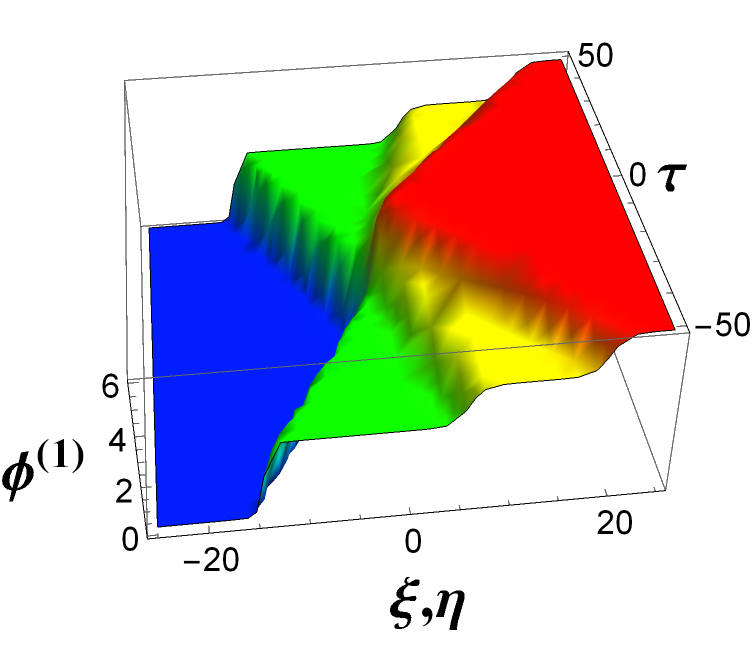}\label{f:2b}}
	\hfill
	\subfigure[]{\includegraphics[width=0.45\linewidth]{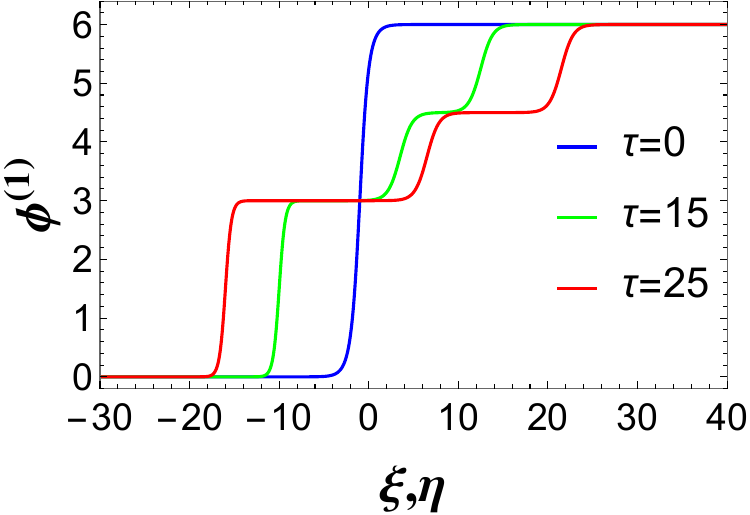}\label{f:2c}}
	\hfill
	\subfigure[]{\includegraphics[width=0.45\linewidth]{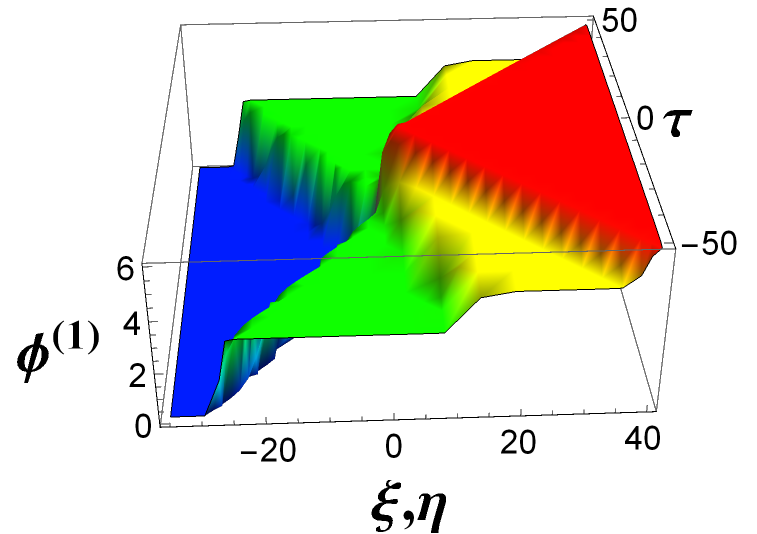}\label{f:2d}}
	\caption{ Collision of two-fold shock wave corresponding to equation (\ref{phi2}) for $\lambda=2$, $\alpha_1=0.1526$, $\alpha_2=0.25$, $\alpha_e=1.344$, $\mu_1=0.628$, $\mu_2=0.5$, $\sigma_1=0.083$, $\sigma_2=0.033=\sigma_3$,$\kappa=3.5$, $\nu_2=0.0$, $\nu_3=0.1$, $k_1=1.3=l_1$ and $k_2=3.3=l_2$, here (a) and(c) are 2D profile for different $\tau$  where $\nu_1=0.2$, and $\nu_1=0.4$ respectively, and (b), (d) are corresponding 3D figure of (a) and (c), respectively.}\label{f:2}	
\end{figure}

\section{Conclusion}\label{sec5}
In this study, we have theoretically investigated the head-on collision dynamics of IA shock waves in a multicomponent Martian ionospheric plasma, consisting of three positive ion species (hydrogen, oxygen, and oxygen molecules) and superthermal electrons in context with observations of MAVEN
spacecraft in Martian atmosphere. Using the hydrodynamic model, the extended PLK perturbation scheme has been employed to obtain a set of Burgers' equations for a pair of shock waves. The study specifically addresses the influence of head-on collisions on the electrostatic potential profiles associated with one-fold and two-fold shock interactions. Our numerical analysis reveals that, following the collision, the spatial width of the interacting shock structures increases with increasing kinematic viscosity, indicating the dispersive-dissipative nature of the plasma environment due to enhanced viscous dissipation. These results provide deeper insight into the nonlinear interaction processes of IA shocks and may contribute to a more comprehensive understanding of wave dynamics in the Martian ionosphere.

\end{document}